\documentclass[twocolumn]{aastex631}

\usepackage{amsmath}
\usepackage{soul}
\usepackage{hyperref}
\newcommand\blfootnote[1]{%
  \begingroup
  \renewcommand\thefootnote{}\footnote{#1}%
  \addtocounter{footnote}{-1}%
  \endgroup
}

\begin{document}

\title{PINN ME: A Physics-Informed Neural Network Framework for Accurate Milne-Eddington Inversions of Solar Magnetic Fields}

\blfootnote{$^*$R. Jarolim and M. E. Molnar contributed equally to this study.}

\correspondingauthor{Robert Jarolim}
\author[0000-0002-9309-2981]{Robert Jarolim$^*$}
\email{rjarolim@ucar.edu}
\affiliation{High Altitude Observatory, 
National Center for Atmospheric Research, Boulder, CO}

\author[0000-0003-0583-0516]{Momchil E. Molnar$^*$}
\affiliation{Southwest Research Institute, Boulder, CO 80301}
\affiliation{High Altitude Observatory, 
National Center for Atmospheric Research, Boulder, CO}

\author[0000-0002-5181-7913]{Benoit Tremblay}
\affiliation{High Altitude Observatory, 
National Center for Atmospheric Research, Boulder, CO}
\affiliation{Environment and Climate Change Canada, Montréal, Québec, Canada}

\author[0000-0000-0000-0000]{Rebecca Centeno}
\affiliation{High Altitude Observatory, 
National Center for Atmospheric Research, Boulder, CO}

\author[0000-0000-0000-0000]{Matthias Rempel}
\affiliation{High Altitude Observatory, 
National Center for Atmospheric Research, Boulder, CO}



\begin{abstract}

Spectropolarimetric inversions of solar observations are fundamental for the estimation of the magnetic field in the solar atmosphere. However, instrumental noise, computational requirements, and varying levels of physical realism make it challenging to derive reliable solar magnetic field estimates. 
In this study, we present a novel approach for spectropolarimetric inversions based on Physics Informed Neural Networks (PINNs) to infer the photospheric magnetic field under the Milne-Eddington approximation (PINN ME). Our model acts as a representation of the parameter space, mapping input coordinates ($t, x, y$) to the respective spectropolarimetric parameters, which are used to synthesize the corresponding stokes profiles. By iteratively sampling coordinate points, synthesizing profiles, and minimizing the deviation from the observed stokes profiles, our method can find the set of Milne-Eddington parameters that best fit the observations. In addition, we directly include the point-spread-function to account for instrumental effects.
We use a predefined parameter space as well as synthetic profiles from a radiative MHD simulation to evaluate the performance of our method and to estimate the impact of instrumental noise. Our results demonstrate that PINN ME achieves an intrinsic spatio-temporal coupling, which can largely mitigate observational noise and provides a memory-efficient inversion even for extended fields-of-view.
Finally, we apply our method to observations and show that our method provides a high spatial coherence and can resolve small-scale features both in strong- and weak-field regions.


\end{abstract}



\section{Introduction}
\label{sec:intro}

\begin{figure*}[!htb]
    \centering
    \includegraphics[width=\linewidth]{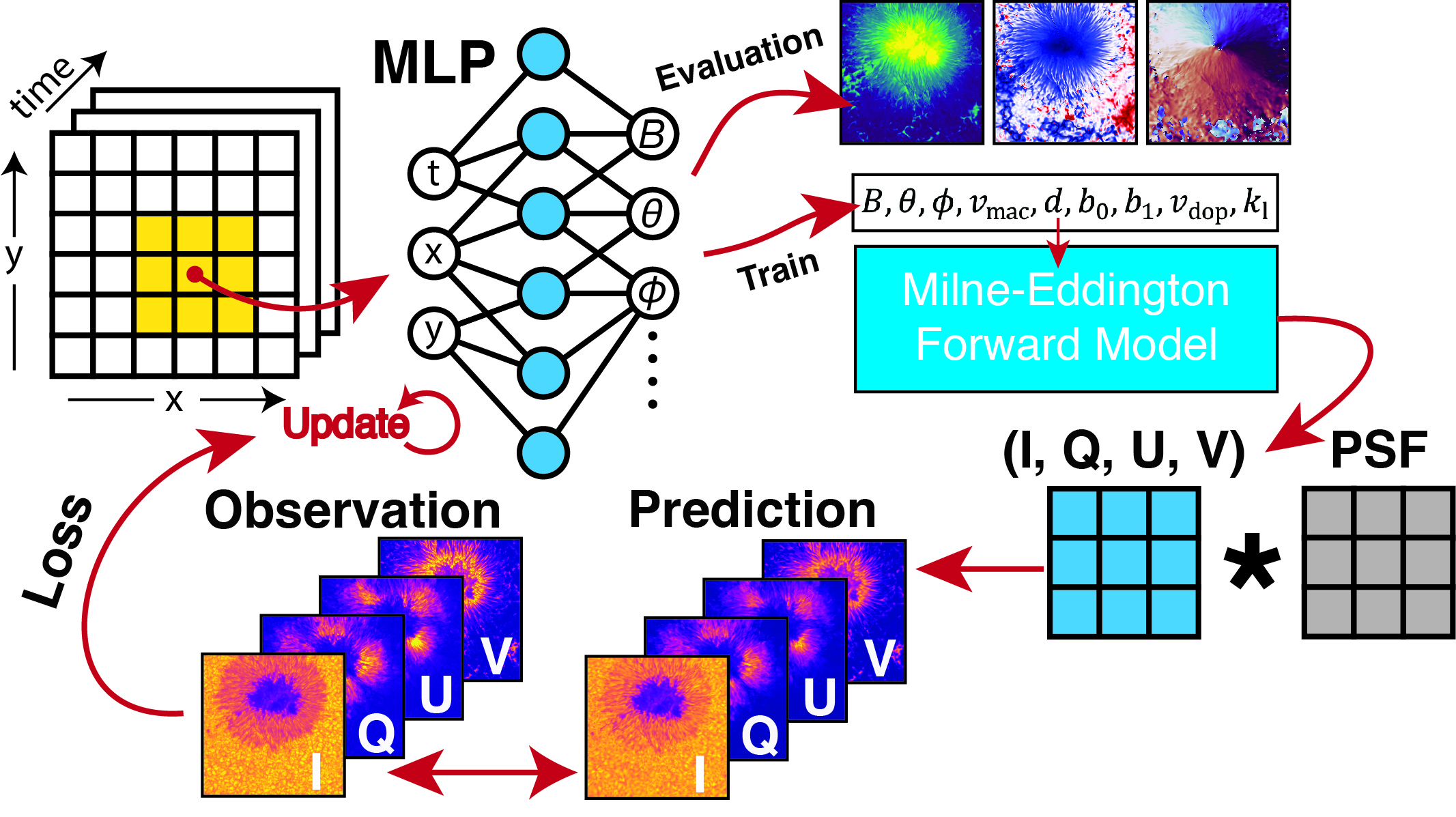}
    \caption{Overview of the PINN ME Milne-Eddington inversion method. The neural network acts as a representation of the parameter space, and maps pixel coordinates ($t, x, y$) of the input image sequence to the respective plasma parameters. For model training, the parameters are used as input to the fully-differentiable ME forward model which yields the corresponding spectral line profiles of the Stokes vector ($I, Q, U, V$). The  difference (loss) between the predicted Stokes vectors and the observed data is used to update the network weights. The model is iteratively updated until the difference between the modeled and observed stokes vectors converges to a minimum. From the trained model, the individual pixel coordinates can be queried to obtain the final inversion result.}
    \label{fig:overview_method}
\end{figure*}

The solar magnetic field is the omnipresent component that shapes the solar atmosphere, the heliosphere, and 
drives space weather~\citep{Owens2013}. The magnetic 
field in the lower solar atmosphere is studied through its 
signatures imprinted on the polarization of light \citep{Casini2017}. For this, spectropolarimetric inversions are used to determine the plasma conditions which most closely
reproduce the observed polarization signal~\citep{delToroIniesta2016}. 

Methods for magnetic field inversions use an iterative approach to fit the observed spectral line properties while employing different levels of sophistication. Recent methods can account for the varying conditions from the photosphere, which is in local thermodynamic equilibrium (LTE), to the chromosphere which is out of LTE conditions \citep{delaCruzRodriguez_2017}. However, computational requirements and numerical instabilities are strong limitations for the application of the most advanced inversions methods to the vast amounts of data current generation of solar telescopes produce~\citep{Reardon_2023}. 

A frequently applied approximation for interpreting lines formed in the solar photosphere is the Milne-Eddington (ME) model atmosphere. The ME approximation poses that the source function changes 
linearly over the formation height 
of the spectral line under consideration, while the other 
atmospheric parameters are constant over that region.
ME inversions are established in the pipelines of many current generation magnetographs, such as the Helioseismic and Magntic Imager \citep[HMI;][]{schou2012hmi} onboard the Solar Dynamics Observatory \citep[SDO;][]{pesnell2012sdo}, the Solar Optical Telescope Spectro-Polarimeter \citep[SOTSP;][]{tsuneta2008sot, Lites2013} onboard Hinode, and the Synoptic Optical Long-term Investigations of the Sun \citep[SOLIS;][]{Keller_2003} operated by the National Solar Observatory (NSO), as they provide the real-time speed required for operational needs. 
Furthermore, their performance has been benchmarked against the different implementations used in the community on realistic radiative magnetohydrodynamic (rMHD) simulations to show high degree
of accuracy \citep{Borrero_2014}. To enable ME inversions to utilize the correlations between the neighboring pixels, inversions with spatio-temporal coupling based on sparse-matrix algorithmic approaches \citep{vanNoort2012} have been used to mitigate observational noise and improve the precision of the obtained magnetic fields \citep{Morosin2020, delaCruzRodriguez2024}. However, these methods typically have large computational memory requirements which are prohibitively challenging for larger field-of-views and longer temporal series (e.g., full-disk observations).

Machine learning methods were applied to solar magnetic field inversions and related applications \citep[see][ for a review]{ramos2023ml}. A frequent approach is the use of simulated data or existing inversion results to perform supervised training \citep{milic2020cnn_inversion, navarro2005nn_inversion, ramos2019inversion, higgins2021nn_inversions, higgins2022synthia, gafeira2021nn_inversion_initialization}. While this approach can achieve significant accelerations of inversion codes, the quality of the resulting inversion strictly depends on the provided data set. In contrast, in this study, we develop a novel inversion method that provides independent and self-consistent solutions, solely based on the observed Stokes vector. This approach sets the foundation to exceed the quality and to overcome challenges of existing inversion methods (e.g., computational demand of non-LTE inversions). 

Physics-Informed Neural Networks \citep[PINNs; ][]{raissi2019pinns} provide a novel approach to smoothly integrate physics-based models and noisy observational data \citep{karniadakis2021pinns}. 
The method uses a neural network to encode the simulation volume, where input coordinates (e.g., $x, y, z$) are mapped to the corresponding physical quantities (e.g., the magnetic field $\vec{B}(x, y, z)$). The model training is performed by fitting a given boundary condition and by minimizing the residuals of the physical model (e.g., partial differential equations. Encoding the simulation volume into weights of a neural network can enable smoother solutions, with an intrinsic coupling of adjacent points. In addition, this encoding is highly memory efficient, making it well suited for high-dimensional problems, and can simultaneously encode small- and large-scale features. In solar physics, the concept of a neural representation was used for magnetic field extrapolations \citep{jarolim2023nf2, Jarolim2024multi, jarolim2024ar13664}, tomographic reconstructions \citep{binsi2022sunerf, jarolim2024sunerf, ramos2023tomography}, and solar flux transport \citep{Athalathil2024flux_transport}. \citet{DiazBaso_2024} used a PINN for spectropolarimetric inversions under the Weak-Field Approximation (WFA) of chromospheric \ion{Ca}{2} 854.2\,nm solar observations.

In this study, we present a Physics Informed Neural Network for Milne-Eddington inversions (PINN ME) and demonstrate that spatio-temporal regularization and the instrumental Point-Spread-Function (PSF) deconvolution can improve magnetic field estimates of noisy data, enabling memory efficient inversions even for extended fields-of-view. We quantify the performance of our method with analytical, synthetic and observational data, and compare to recent state-of-the-art inversion methods (Sect. \ref{sec:results}).

\section{Method}
\label{sec:method}

\begin{figure*}[t]
    \centering
    \includegraphics[width=\linewidth]{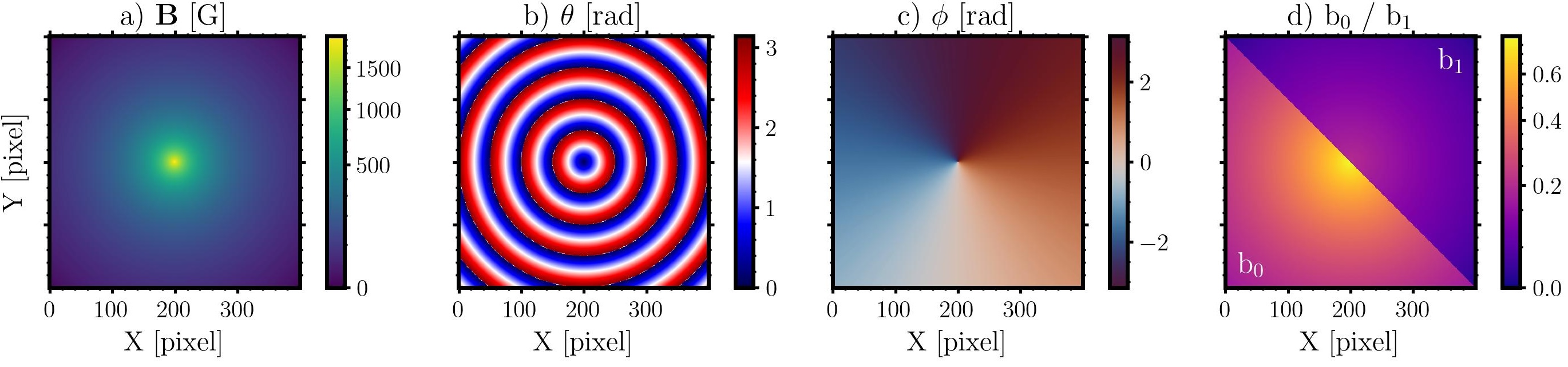}
    \caption{Overview of the synthetic dataset used for testing and benchmarking. a) Magnetic field strength (gauss); b) magnetic field inclination (radians); c) magnetic field azimuth (radians); d) the constant/gradient parameters 
    of the source function $b_0$/$b_1$ in the left/right part of the panel.}
    \label{fig:overview_testset}
\end{figure*}

Our PINN ME method performs the spectropolarimetric inversion of the full Stokes vector $I,Q,U,V$ under the Milne-Eddington assumption to obtain vector magnetic field estimates from solar observations ($\vec{B}$). Here, the neural network maps pixel coordinates of the input Stokes profile to the corresponding ME parameters ($B, \theta, \phi, v_\text{mac}, d, b_0, b_1, v_{\text{dop}}, k_l$; see Sect. \ref{sec:method_me}). The ME parameters are used to synthesize the corresponding Stokes profile, by using a fully-differentiable ME forward model (Sect. \ref{sec:method_me}).
The resulting $I, Q, U, V$ predictions are compared to the observed 
Stokes profile, from which we compute the loss (Sect. \ref{sec:mehtod_pinn}). By iteratively minimizing the difference between the observation and the predicted Stokes profiles, the full set of ME parameters is found, similar to pixel-wise ME inversion methods \citep[e.g.,][]{Lites_2007, Hoeksema2014hmi}. In contrast to classical ME inversion methods, this approach provides an intrinsic coupling between adjacent pixels and optimizes the full frame simultaneously. This intrinsically provides smooth solutions which can mitigate observational noise. Furthermore, this encoding is memory efficient, which allows to perform spatio-temporal coupled inversions over extended time sequences, at full resolution. We further account for the instrumental PSF by sampling coordinate points according to the PSF grid (e.g., $3\times3$). The resulting Stokes profiles are then convolved with the PSF to obtain the corresponding pixel value of the predicted Stokes profile. This further improves the spatial coupling and directly enables a deconvolution of the deduced parameter space. Figure \ref{fig:overview_method} provides an overview of the end-to-end training procedure.
Note that no strict boundary condition is applied (e.g., periodic boundary on the side boundaries, padding). Therefore, sampling points at the boundary are only partially constrained by the adjacent observational data, and can be chosen arbitrarily by the neural network. Consequently, inversion results close to the boundary are expected to be less reliable.

\subsection{Milne-Eddington forward model}
\label{sec:method_me}

The Milne-Eddington (ME) model assumes a
linearly changing source function $S(\tau)$ with respect to the optical depth $\tau$ in the stellar atmosphere, where the ratio of continuum to line absorption coefficients is constant
\citep{Unno_1956,Rachkovsky_1962,Rachkovsky_1967}
\begin{equation}
    S(\tau) = b_0  - b_1 \tau \,,
\end{equation}
where $b_0$ and $b_1$ are coefficients that determine the overall intensity and line depth.
The other atmospheric parameters -- magnetic field strength $B$, inclination angle $\theta$, azimuth angle $\phi$, Doppler velocity $v_\text{dop}$, macroturbulent velocity $v_\text{mac}$, line dampening $a_l$, and the 
continuum to line absorption coefficient ratio $k_l$ -- 
are constant valued (i.e. they have no gradients). The ME approximation holds well for spectral lines formed over a 
narrow region of the atmosphere, compared to the local atmospheric scale height, resulting in a slowly changing source function~\citep{1982SoPh...78..355L}.
Also, the ME approximation does not work well for cases with strong velocity gradients, as it is unable to reproduce asymmetric profiles. Based on this, the ME approximation can be valid for spectral lines formed in the photosphere, and for optically thin lines formed higher in thin slabs such as the \ion{He}{1} 1083 nm line.

The advantage of the ME approximation is its fully analytical description of the 
emergent polarized intensity given a set of input parameters. The simplicity of its fully analytical expressions allows for fast numerical implementation 
as well as analytical uncertainty estimates
\citep{OrozcoSuarez2007}. Nowadays, ME inversions are the foundation of most photospheric magnetic field studies. 

\subsection{PINN setup}
\label{sec:mehtod_pinn}

PINN ME is a fully-connected neural network with eight hidden layers, each consisting of 256 neurons with sine activation functions \citep[i.e., SIREN][]{sitzmann2020siren}. The input layer uses the three coordinate positions ($x, y, t$) and the output layer provides the nine ME parameters ($B, \theta, \phi, v_\text{mac}, d, b_0, b_1, v_{\text{dop}}, k_l$). We employ positional encoding to map the input coordinates into Fourier features \citep{tancik2020fourier}. For this, we select 20 random frequencies per coordinate from a normal distribution and encode the input coordinates 
\begin{align}
    \vec{v}_{\rm encoded} = \big[
        &\sin{(f_{x, 0} \cdot x)}, \cos{(f_{x, 0} \cdot x)}, \sin{(f_{x, 1} \cdot x)}, \ldots \notag \\
        &\ldots, \sin{(f_{t, 19} \cdot t)}, \cos{(f_{t, 19} \cdot t)},  x, y, t
    \big] \,,
\end{align}
where $f_{i,j}$ refers to the $j$th sampled frequency of coordinate $i$.
This enables the neural network to efficiently learn high frequency features. The model provides a total of 560,000 free parameters to encode the full parameter space. The input coordinates are normalized to range approximately between 0 and 1, with the same grid spacing in the spatial dimensions. For the time coordinate, we set the first observation as the reference point (time $t = 0$), calculate the time differences relative to it, and scale the values such that one hour corresponds to a normalized value of 1. For inversions of a single frame the time coordinate is set to 0. For the output parameters, we rescale the model outputs to match the expected value range. For the magnetic field strength ($B$), we use a linear output and rescale it by a factor $10^3$. Both the inclination $\theta$ and the azimuth angle ($\phi$) use linear outputs and are scaled by $\pi$. Therefore, we do not restrict the value range of the angles, but reproject them to [0, $\pi$] after the model training. For the rest of the parameters -- macroturbulent velocity $v_\text{mac}$, the line damping $d$, the Doppler velocity $v_{\text{dop}}$, the ratio of continuum to line absorption coefficient $k_{l}$, and the source function intercept and slope ($b_0$, $b_1$), we use a fixed value range by applying a sigmoid output activation (value range of [0, 1]) and scale the parameters by $2\times 10^4$, $1.0$, $10^4$, $10^2$, $1.0$, and $1.0$, respectively.

As optimization target we use the observed Stokes vectors. We apply an additional weighting and scaling to increase the fitting of weak field regions and to enable a balanced optimization of the four Stokes parameters. We normalize the individual Stokes parameters by the maximum value in the data, such that $I$ ranges between $[0,1]$, and $Q,U,V$ range between $[-1,1]$. The resulting Stokes vector is normalized through an asinh stretch
\begin{equation}
    \vec{S}_\text{scaled} = \frac{\text{asinh}(\vec{S} \cdot 10^{2})}{\text{asinh}(10^{2})} \,,
\end{equation}
where $\vec{S}$ refers to the Stokes vector. This provides a linear scaling close to 0 and a logarithmic scaling for larger values, resulting in a better balancing between the optimization of weak and strong magnetic field regions.

\subsection{Reference method - PyMilne}

We compare the results from PINN ME to 
PyMilne, a state-of-the-art ME inversion method \citep[][]{delaCruzRodriguez2019}. PyMilne uses an efficient hybrid implementation in Python/Cython, which allows for spatially-coupled inversions including the instrumental PSF and spatio-temporal regularization \citep{Morosin2020,delaCruzRodriguez2024}. PyMilne uses a standard sparse matrix approach for solving the coupled equations system \citep{delaCruzRodriguez2019}. This provides improved magnetic field inversions, but has significant memory requirements. Therefore, this method provides an ideal reference to assess the performance of our inversions and to outline the advantages of the PINN approach in terms of memory consumption. We note that we use very similar atomic parameters to the one in PyMilne, based on the NIST atomic database \citep{NIST_ASD}.

\subsection{Data}
\label{sec:method_data}
In this study, we consider three data sets to validate our approach and to compare to recent state-of-the-art inversion methods.

\textbf{(1) Analytical data:} We synthesize Stokes profiles from a predefined parameter space. We use this data set to validate the self consistency of the method against 
other ME inversion methods. Using synthetic 
profiles from ME synthesis to benchmark
the performance of the method in its range of validity, compared to using more
complicated radiative transfer products tests only the performance of the code
\citep{Borrero_2014}.

The atmospheric parameters in this dataset have either an azimuthal or radial symmetry, as shown in Figure~\ref{fig:overview_testset}. We also include sharp discontinuities in the magnetic field inclination (panel b in Figure~\ref{fig:overview_testset}) paired with a wide range of field strengths to test the accuracy and precision of the inversion code under different magnetic field strength regimes, while also probing the extent to which PINN ME can recover the discontinuities when a spatial PSF is applied and accounted for. We also extend the dataset in time by increasing the length scales of the concentric circles (video available online), to test the temporal regularization of our inversion approach. We define the magnetic field strength $B$ as
\begin{equation}
    B = B_0 \cdot \left(\frac{r_0}{r + r_0}\right)^2 \,,
\end{equation}
the inclination angle $\theta$ in radians as
\begin{equation}
    \theta = \frac{r \mod r_0}{r_0} \cdot\pi \,,
\end{equation}
the azimuth angle $\phi$ in radians as
\begin{equation}
    \phi = \arctan{(\frac{y}{x}}) + \frac{t}{180} \cdot \pi \,,
\end{equation}
and the ME parameters $b_0$ and $b_1$ as
\begin{equation}
    b_{0,1} = a_{0, 1} \left(\frac{10 \cdot r_0}{r + 10 \cdot r_0}\right)^2\,,
\end{equation}
where $t$ refers to the timestep (20 frames; [0, 19]), $x$ and $y$ to the spatial axis ([-200, 200]), $r$ to the radius ($r = \sqrt{x^2 + y^2}$), $B_0$ to the scaling of the magnetic field strength ($B_0 = 2000$), $r_o$ to the scaling factor ($r_0 = 50 + \frac{t}{2}$), and $a_{0,1}$ to the scaling factor of the ME parameters, which we set to $a_0 = 0.8$ and $a_1 = 0.2$. The remaining four parameters are set constant with $v_\text{mac}=2.0e3$, $d=0.2$,  $v_\text{dop}=2.0e3$, and $k_l=25$.
We synthesize the \ion{Fe}{1} 630.2 nm line with the PyMilne code~\citep{delaCruzRodriguez2019}, where we
utilize the wavelength sampling parameters of the Hinode/SOTSP telescope 
(wavelength spacing of 0.0217\,{\AA}). 


The obtained profiles are further degraded to quantify the impact of instrumental effects (i.e., noise, PSF) on our inversion method. For this we construct a $3\times3$ PSF from a normal distribution with $\sigma=1$. For each Stokes vector, and spectral point, we convolve the image with the PSF and a normal distributed random noise factor per pixel. Here, we vary the level of the noise between zero (only the PSF), 1e-4, 1e-3, and 1e-2.

\textbf{(2) Simulated data:} We use synthetic Stokes profiles of the \ion{Fe}{1} 630.2 nm line which were synthesized from a MURaM radiative MHD simulation \citep{voegler2005muram, rempel2012} of a sunspot, used previously in \citet{Borrero_2019}. The simulation box size is $4096\times4096\times768$ in $(x,y,z)$ with a grid size of $\Delta$x = 12, $\Delta$y = 12, and $\Delta$z = 8 km. As in \citet{Borrero_2019}, we used a subset of the domain spanning of size $4096\times512\times192$ grid points covering $49.152\times6.144\times1.536$ Mm$^3$, centered on the sunspot.Here, the full 3D magnetic field was obtained from the  simulation, and the synthetic observables were synthesized using the Stokes Inversion based on Response functions code \citep[SIR; ][]{RuizCobo_1992}. Note that this method synthesizes line profiles under the LTE assumption, resulting in more complex profiles than could be fitted by an ME approach. We further reduce the resolution by a factor 4, apply the PSF from Hinode/SOTSP, and add a noise factor of 1e-3 to the profiles (analogously to (1)). From this we obtain synthetic observations with a high degree of realism. By comparing the inversion result of magnetic field parameters to the known ground-truth field we can estimate the model performance in a realistic setting. While the LTE profiles provide a more realistic approximation of solar observations, the parameters cannot be mapped directly to a specific height surface. 


\textbf{(3) Observational data:} We use observations from Hinode/SOTSP to demonstrate the applicability of our approach and to compare to the conventional pipeline inversions. The observations are obtained from a slit spectrograph. The considered observation has a spatial resolution of [0.15, 0.16] arcsec per pixel, and a spatial extent of $153.6\times81.92$~arcsec$^2$. The observations were obtained from 2007 January 5th 23:59:07 to 6th 01:25:57, close to the disk center, and include both strong and weak field regions (see Fig. \ref{fig:hinode_comparison}). For comparison against the PINN ME results, we use the standard pipeline inversions\footnote{\url{https://csac.hao.ucar.edu/sp_data.php}} provided by the Milne-Eddington gRid Linear Inversion Network \citep[MERLIN;][]{Lites_2007} code. Here, we adjust the outputs from MERLIN to account for the filling factor correction
($\overline{B}_\text{LOS}=s \cdot B_\text{LOS}$, 
$\overline{B}_\text{TRV}= \sqrt{s} \cdot B_\text{TRV}$, where $s$ corresponds to the stray light correction factor).

\begin{figure}[t]
    \centering
    \includegraphics[width=0.74\linewidth]{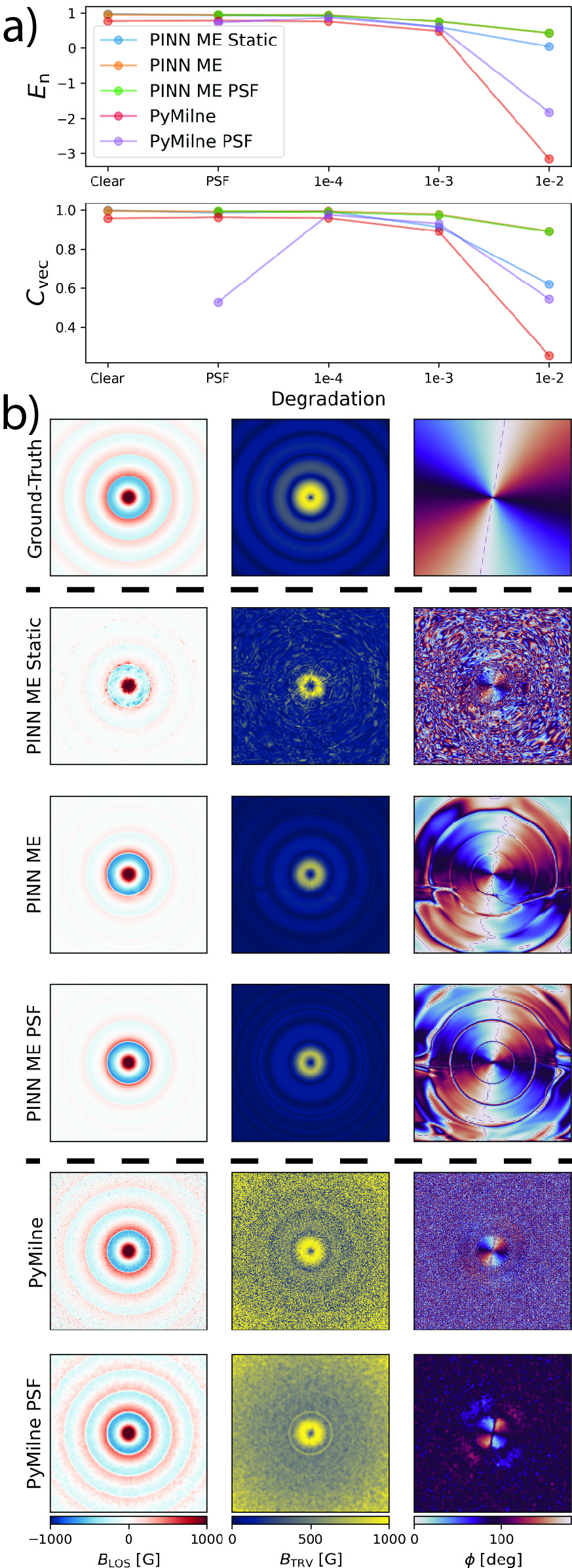}
    \caption{Comparison of PINN ME and PyMilne performance for the test set under different noise levels. a) Model performance as function of data degradation. We consider the unmodified ground-truth data (clear), the PSF-convolved data (PSF), and PSF-convolved data with added noise levels of $10^{-4}$ to $10^{-2}$. b) Inferred magnetic field parameters of the individual models for the PSF-convolved $10^-2$ noise case. The top panel shows the Ground-Truth magnetic parameters ($B_\text{LOS},  B_\text{TRV}, \phi$). We show the inversion results from PINN ME, with the static assumption, the included time component, and the additional PSF, as well as PyMilne inversion results for the pixel-wise inversions, and the spatially coupled inversions with included PSF.
    }
    \label{fig:evaluation_test_set}
\end{figure}

\begin{figure*}[ht]
    \centering
    \includegraphics[width=\linewidth]{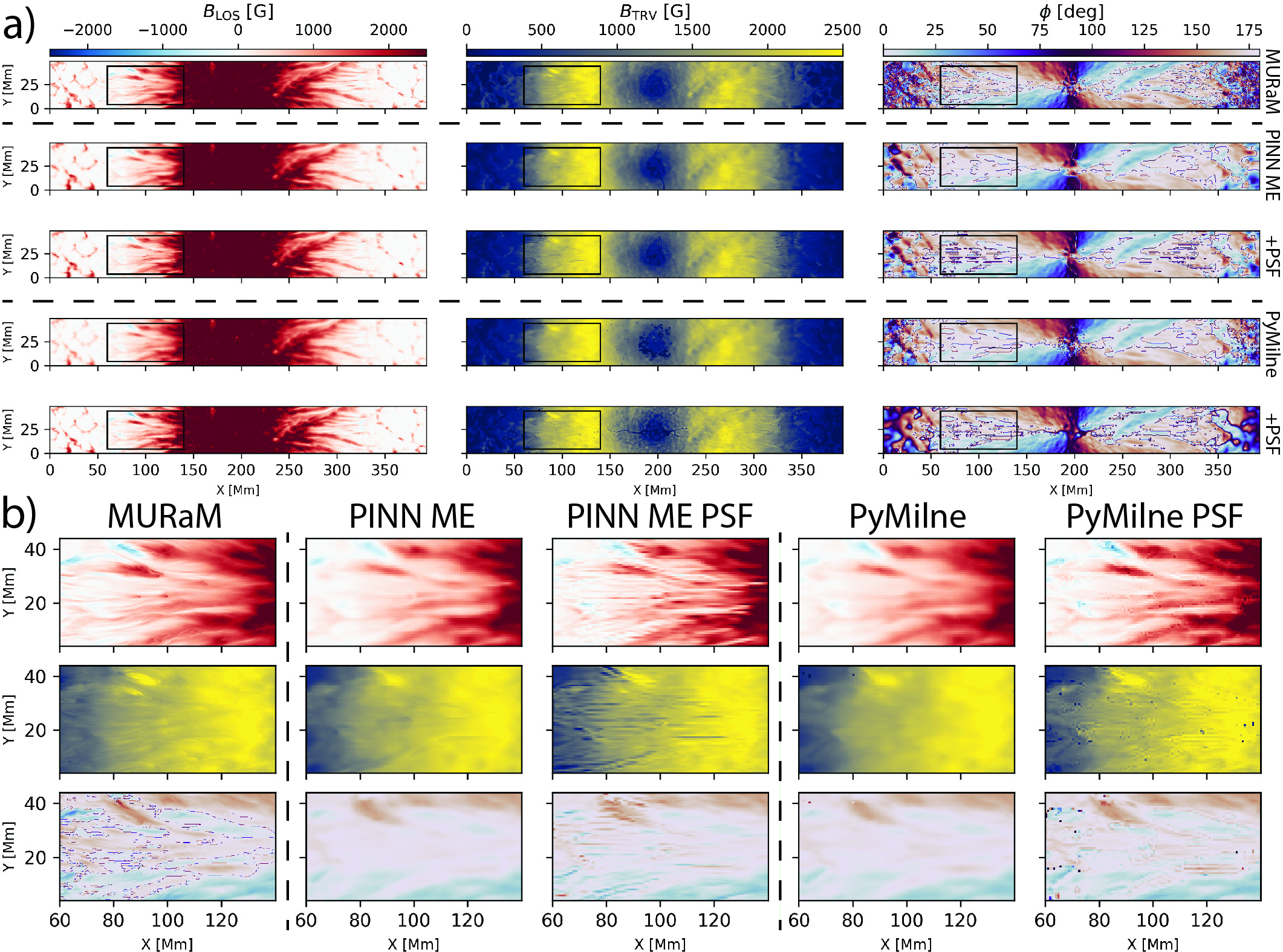}
    \caption{Results of PINN ME and PyMilne inversions of the synthetic stokes profiles from the MURaM simulation. a) Magnetic field LOS component (left column), transverse component (center column), and the magnetic field azimuth (right column) for the considered model configuration (labeled on the right). The ground-truth (MURaM) field corresponds to the $\tau=10^{-2}$ layer. b) Zoom in of the region of interest outlined by the black rectangle in panel a for the different models.
    }
    \label{fig:muram_comparison}
\end{figure*}

\subsection{Evaluation Metrics}

While ME inversions provide nine free parameters, we focus our evaluation solely on the derived magnetic field. Milne-Eddington inversions provide the absolute magnetic field strength $B$, the inclination angle of the field $\theta$, and the azimuth angle $\phi$, from which we compute the line-of-sight magnetic field ($B_{\text{LOS}}=B\cdot\cos{\theta}$) and the transverse field $B_{\text{TRV}}=B\cdot\sin{\theta}$). 
In addition, we reproject the azimuth angle to a value range of $[0, \pi)$, to exclude the 180 degree ambiguity from our evaluation. From the full magnetic field vector ($\vec{B}=(B_{\text{LOS}}, B_{\text{TRV}}\cdot\cos(\phi), B_{\text{TRV}}\cdot\sin(\phi))$), we compute normalized error metrics between the reference magnetic field $\vec{B}_{\rm ref}$ and the inversion result $\vec{B}$ \citep{schrijver2006NLFFcomparison}.
To measure the difference between vectors, we compute the mean error normalized by the average vector norm
\begin{equation}
    E_\text{n} (\vec{B}_{\rm ref}, \vec{B}) = \frac{\sum_i \lVert \vec{B}_{i} - \vec{B}_{\text{ref}, i} \rVert}{\sum_i \lVert \vec{B}_{\text{ref}, i} \rVert} \,.
\end{equation}
Here, $i$ denotes the index of the grid cell, and the sum is computed over all grid cells. The best performance is achieved when $E_\text{n} = 0$. We set $E'_\text{n} = 1 - E_\text{n}$, such that the best attainable performance corresponds to 1. In addition, we use the vector correlation coefficient ($C_\text{vec}$), which compares the local characteristics of the magnetic field vectors
\begin{equation}
    C_\text{vec}(\vec{B}_{\rm ref}, \vec{B}) = \frac{\sum_i \vec{B}_{\text{ref}, i} \cdot \vec{B}_{i}}{\left(\sum_i \lVert \vec{B}_{\text{ref}, i} \rVert^2 \sum_i \lVert \vec{B}_{i} \rVert^2\right)^{1/2}} \,.
\end{equation}

\section{Results}
\label{sec:results}

\begin{figure*}[ht]
    \centering
    \includegraphics[width=\linewidth]{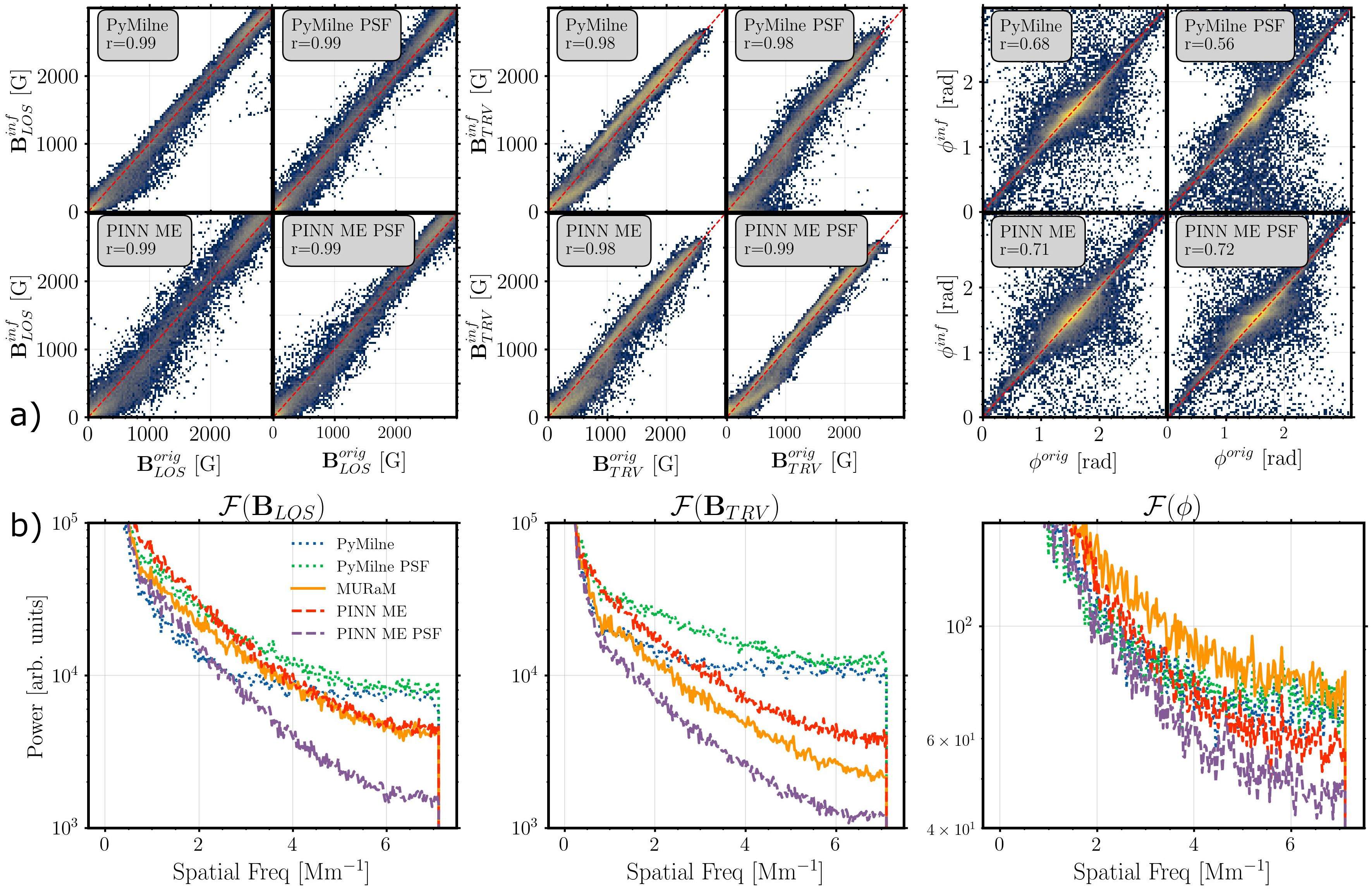}
    \caption{Evaluation of the inversion results of the synthetic MURaM stokes profiles. a) Histograms between the inferred and true line-of-sight
    magnetic fields for the four considered model configurations (PyMilne, PyMilne PSF, PINN ME, and
    PINN ME PSF). The red lines represents the ideal correspondence between the reference and inverted parameters. The left most panel shows the LOS magnetic field histograms, the middle one the transverse magnetic field, and the right one the magnetic field azimuth. Panels b) show the power spectra of 
    the LOS magnetic field (left panel), transverse magnetic field (middle panel), and the magnetic 
    azimuth angle (right panel). 
    }
    \label{fig:MURaM_inversion_props}
\end{figure*}

In this section, we perform a quantitative evaluation of our method with synthetic Stokes profiles and provide a qualitative comparison with observational data. 
Throughout this section, we compute $E_\text{n}'$ and $C_\text{vec}$ for the quantitative evaluation and use the line-of-sight magnetic field strength $B_{\text{LOS}}$, the transverse magnetic field strength $B_{\text{TRV}}$, and the azimuth angle $\phi$ for visualizations.

\subsection{Test set}
\label{sec:results_testset}

We use the analytical test set to validate the self-consistency of our method. The sharp transitions in the inclination angle and the drop off in polarization signal from the center make this a challenging application for inversion methods (see Fig. \ref{fig:overview_testset}).  We apply both our PINN ME and the PyMilne inversion methods to the test set where we vary between the unmodified (clear) profiles, the PSF-convolved profiles, and the PSF-convolved profiles with three noise levels ($1e-4$, $1e-3$, $1e-2$). In Fig. \ref{fig:evaluation_test_set}, we show the evaluation of the derived parameters. We compare our PINN ME approach, with and without the additional PSF sampling, and PyMilne inversions using regular pixel-by-pixel inversions and spatially regularized inversions including the spatial PSF (PyMilne PSF). The PyMilne inversions are applied to a single frame (frame 10), while our PINN ME inversions are applied to the full time sequence. For reference, we also perform inversions with PINN ME for a single time step (PINN ME static), which also accounts for the PSF. All evaluations are performed with the central time step (frame 10) of the sequence. The quantitative evaluation shows the clear trend of degrading inversions with increasing noise level ($E_\text{n}'$, panel a). The derived solutions with and without PSF show almost identical solutions for PINN ME. For PyMilne the additional inclusion of the PSF provides an improvement over the pixel-wise inversions. From $C_\text{vec}$ it can be seen that PINN ME achieves throughout a better estimation of the magnetic field parameters, where the biggest improvement is achieved by including the temporal evolution. Especially in the case of strong noise, the performance decrease of PINN ME is small compared to PyMilne inversions. This is largely attributed to the temporal regularization, as can be seen from the larger performance decrease of PINN ME Static. Panel b shows the obtained magnetic field parameters for the strong noise case ($10^{-2}$). The comparison illustrates that the weak field regions, with a low signal-to-noise ratio, dominate the error. Here, PINN ME and PyMilne show equal amounts of noisy reconstructions in the low signal region, while the temporal inversions largely improve the results and recover valid magnetic field orientations and strengths. PyMilne PSF shows a distinct performance decrease of $C_\text{vec}$ for the PSF convolved profiles. This is a result of the spatial regularization close to the sharp change of the inclination angle, where the spatially coupled inversions result in a larger deviation at the discontinuity. Note that both the spatial and temporal smoothness of PINN ME are not explicitly enforced, but are a result of the neural encoding of the parameter space, which favor coherent solutions~\citep[c.f.,][]{DiazBaso_2024}. Therefore, our method provides spatial and temporal smoothness, while also correctly accounting for sharp discontinuities.

\begin{figure*}[ht]
    \centering
    \includegraphics[width=0.9\linewidth]{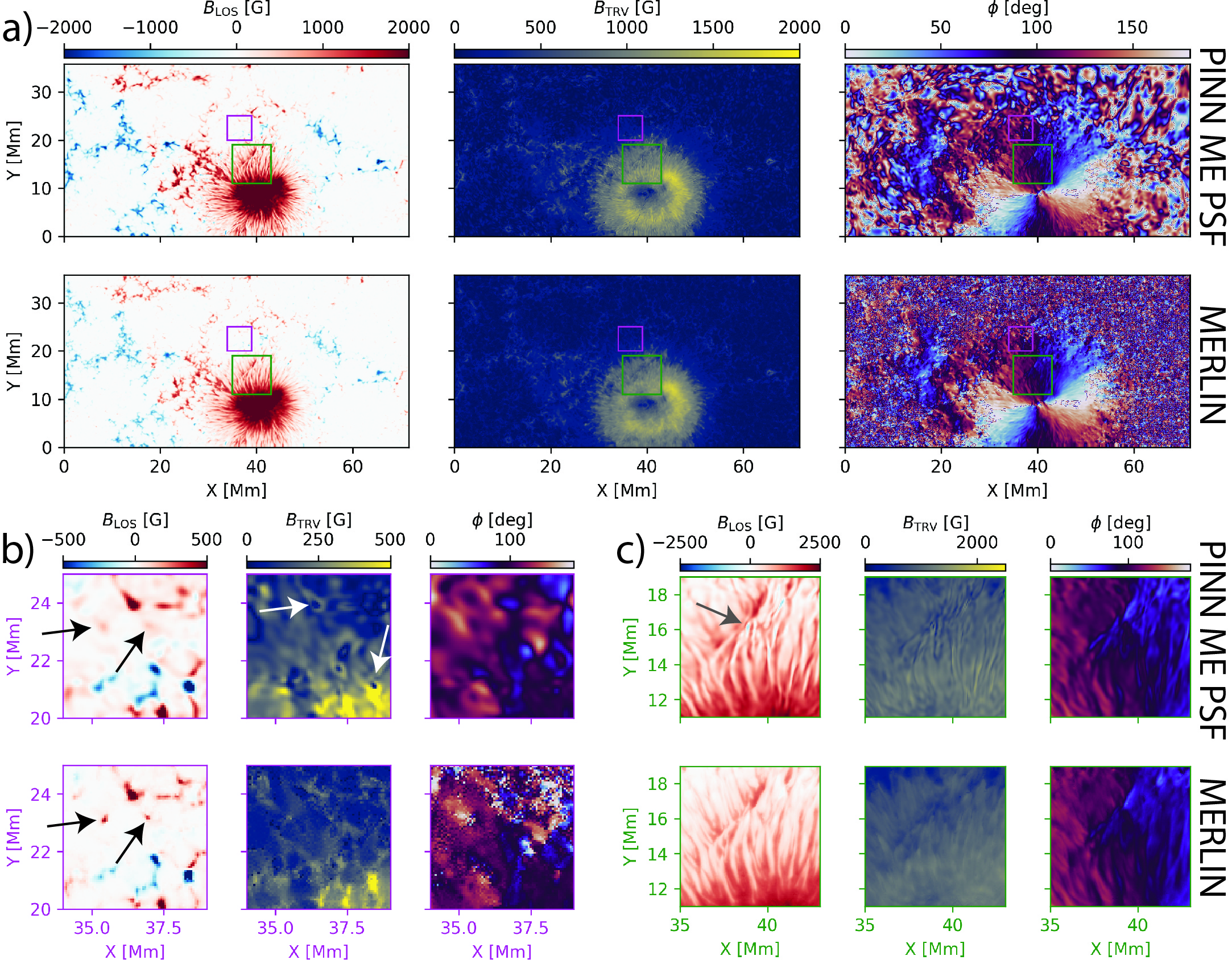}
    \caption{Inversion results from PINN ME PSF and
             MERLIN for Hinode/SOTSP observations of NOAA AR 10933. a) Full field-of-view 
             results for the inferred LOS magnetic field, transverse
             magnetic field, and the magnetic field azimuth angle. b) Zoomed-in quite-Sun region as indicated by the magenta square in panel a. c) Zoomed-in region of the sunspot penumbra as indicated by the green square in panel a.}
    \label{fig:hinode_comparison}
\end{figure*}

\subsection{Synthetic profiles from simulation}

The test case illustrates the validity and applicability of PINN ME for an idealized inversion problem, and demonstrates that limitations imposed by noise and the PSF can be largely mitigated by the concept of a neural representation. The test set represents a smooth parameter space and includes sharp discontinuities. However, to estimate the applicability to solar observations, the additional structural complexity and more realistic spectral line profiles need to be considered. In this section, we use synthetic line profiles obtained from a MURaM simulation of a sunspot to estimate the performance of PINN ME in a more realistic setting that is closer to actual observational data. Here, we use spectral lines that are synthesized under LTE conditions, resample the data to the Hinode/SOTSP resolution, and apply the instrumental PSF and add $10^{-3}$ noise to mimic solar observations.

Figure \ref{fig:muram_comparison} shows the result of PINN ME and PyMilne inversions, both with and without the PSF correction. Overall, both methods show a good agreement with the reference MURaM data, especially on the large scales. The subframe in panel b provides a closer comparison of the sunspot penumbra. Here, both PINN ME and PyMilne achieve a similar degree of spatial resolution, which is slightly smoother than the reference field. By including the PSF, both methods show a strong increase in resolution, however the enhanced features only show a partial agreement with the MURaM reference, and the contrast of $B_\text{TRV}$ is artificially high. For PyMilne, the addition of the PSF further leads to small scale artifacts. 

In Figure \ref{fig:MURaM_inversion_props}, we provide a quantitative comparison of the results, where we plot 2D histograms of the true and inferred $B_\text{LOS}$, $B_\text{TRV}$, and the azimuth angle. The overall scatter of $B_\text{LOS}$ and $B_\text{TRV}$ is low, and show a good one-to-one correspondence (correlation coefficient $r=0.99$). For the azimuth angle, larger deviations from the reference field are present for both methods, where PINN ME achieves a better correlation coefficient with $r=0.72$, while PyMilne achieves $r=0.68$ and $r=56$ for the case of spatially coupled and pixel-wise inversions, respectively. In panel b, we present the power spectra of the two dimensional maps of the inferred physical parameters, which were computed from the amplitude of the two dimensional Fourier transforms of the physical parameters, averaged in the azimuthal plane. The power spectra show that PINN ME retrieves more small scale features compared to the PyMilne inversions, given by the white noise floor that is being reached at higher frequencies. Furthermore, most notable, the white noise floor in the power spectra of the PINN ME inversions are reached at lower values, as shown in Figure~\ref{fig:MURaM_inversion_props}. The lower white noise floor demonstrates that the PINN ME inversions have on average lesser white noise in them, compared to the reference methods exhibiting higher white noise floors in the power spectra. However, note that in some cases PINN ME inversions have significantly lower white noise limit compared to the MURaM case, which is related to the encoding of the neural field in our implementation. Note that the spectral profiles are not obtained from a single height surface, therefore it is not possible to uniquely compare the derived parameters. Here, we compare to the magnetic field parameters at the primary formation height at an optical depth of $\tau=1e-2$. Note that the inversion without the additional PSF results in a smoother solution, and consequently a better agreement with magnetic parameters at a lower optical depth is expected.

\subsection{Application to Hinode/SOTSP observation}

In this section, we apply our inversion method to observations from Hinode/SOTSP and provide a qualitative comparison to the magnetic field that is provided by the standard Hinode/SOTSP pipeline (MERLIN). Figure \ref{fig:hinode_comparison} shows the magnetic field components ($B_\text{LOS}$, $B_\text{TRV}$, and the azimuth angle) as obtained from our inversion including the PSF correction and the MERLIN inversions including the correction for the filling factor described in Section~\ref{sec:method_data}. The full field of view indicates that both methods recover a similar magnetic field distribution (panel a), however the PINN ME inversions lead to slightly stronger magnetic fields and to smoother distributions of the azimuth angle in the quiet-Sun region. 

Panel b shows a subframe of the a quite-Sun region. Here our inversions show a smoother distribution of magnetic elements, as compared to the MERLIN inversions. In particular, the two positive magnetic elements (black arrows) are recovered as more extended regions in our inversions. $B_\text{TRV}$ of the MERLIN inversions show a lower spatial coherence than our PINN ME inversions. A distinct difference is that stronger magnetic elements ($|B_\text{LOS}|\ge500$G) show a vertical orientation in the PINN ME inversions (white arrows), while the MERLIN inversions show a horizontal field component. The azimuth angle shows the largest difference in terms of spatial smoothness, where the PINN ME inversions appear more spatially correlated.

Panel c provides the subframe of the sunspot penumbra, noted as the green square in panels a. In contrast to the quite-Sun region, PINN ME inversions show a higher contrast in the inferred parameters. In particular, $B_\text{TRV}$ shows a clearer separation of individual fibrils, while retrieving stronger horizontal magnetic fields. From $B_{LOS}$ it can be seen that PINN ME resolves small scale negative polarity elements (gray arrow) that are not present in the MERLIN inversions. While there is no ground-truth available to estimate the validity of the inferred magnetic fields, the small scale opposite polarity elements in the penumbra are similar to features in the MURAM simulation (Fig. \ref{fig:muram_comparison}), where also negative polarities are present in the positive magnetic field of the sunspot penumbra.

p\section{Discussion}
\label{sec:discussion}

In this study, we have developed a novel method for magnetic field inversions under the Milne-Eddington approximation. We build on PINNs to encode the space of ME parameters into the weights of a neural network. For this, the neural network maps coordinate points ($x, y, t$) to the respective inversion parameters ($B$, $\theta$, $\phi$, etc.). This encoding enables an intrinsic spatial and temporal coupling of observations, which largely mitigates instrumental noise and limitations of the ME assumption \citep[c.f.][]{DiazBaso_2024}. In addition, we propose to perform a spatial sampling of the PSF which enables an intrinsic deconvolution. Our method has no limitations in memory requirement, and can be applied to datasets with large fields-of-view and temporal extent, expanding current 
capabilities limited by memory constraints.

We perform a thorough evaluation, where we consider an idealized test set, which we synthesize from a pre-defined complex parameter space (Fig. \ref{fig:overview_testset}); synthetic spectral line profiles obtained from a realistic full MHD simulation and synthesized under LTE conditions with the SIR code; and observational data from Hinode/SOTSP. The application to the test set demonstrates that the intrinsic spatial and temporal regularization can achieve valid inversions even in the case of low signal-to-noise. Here, our PINN ME approach substantially outperforms pixel-by-pixel inversions. The indirect spatial regularization prevents artificial smoothing close to sharp transitions as can be seen by the comparison to PyMilne inversions, which uses an explicit regularization scheme. The primary improvement of the inversion is achieved by including the temporal evolution, where even in the case of a changing field, the temporal information provides the decisive component to recover signal in case of high noise (Sect. \ref{sec:results_testset}). The application to synthetic profiles from a MURaM simulation demonstrates that our method is applicable to complex data that is similar to solar observations, and can account for spectral lines that are not in agreement with the ME assumption. The comparison to PyMilne demonstrates that similar results are obtained. Overall, inversions that explicitly model the PSF lead to a higher contrast of the obtained magnetic field and better resolved small scale features, however a direct comparison is limited since the synthetic line profiles cannot be directly associated with a distinct height surface, as assumed by the ME inversions. The application to observational data demonstrates that our inversions provide a better spatial coherence than the regular Hinode/SOTSP data product (MERLIN inversions). In the quiet-Sun and plage regions, our method infers more extended magnetic elements as compared to the MERLIN inversions, while for the strong magnetic field of the sunspot penumbra, individual fibrils are resolved with higher spatial resolution. 
The two most prominent differences between PINN ME and MERLIN inversion are the opposite polarity elements that are resolved in the sunspot penumbra by our approach, while they are not present in the MERLIN inversions; and 
the stronger horizontal fields inferred from PINN ME compared to MERLIN. Comparison with simulated MURaM data suggests that these features are physically sound.

Our PINN ME model is used to invert Stokes profiles over large fields-of-view and time sequences. For the synthetic MURaM sunspot, the inversion process converges in approximately two hours using four V100 GPUs. The primary advantage of the PINN approach is its efficient encoding of patterns in the data, enabling similar convergence times even for larger data volumes \citep[cf. neural network-based tomographic reconstructions;][]{jarolim2024sunerf}. A potential future application is the continuous inversion of SDO/HMI data, where a running sequence of frames could be inverted with minimal additional computational cost \citep[c.f., PINN based force-free extrapolations ][]{jarolim2023nf2}. An alternative approach to improving computation times could involve using pre-trained states, which may facilitate faster convergence \citep[e.g., meta-training algorithms][]{nichol2018reptile}. However, the primary motivation of this study is an initial demonstration of PINNs for spectropolarimetric magnetic field inversions. Here, our approach demonstrates that we achieve a similar performance as recent state-of-the-art methods. The distinct advantage is the ability to solve high-dimensional problems in a memory efficient way and to mitigate observational noise. This provides the foundation and demonstrates the large potential to apply PINNs for LTE and non-LTE inversions, where the neural network can efficiently encode extended solar atmospheres ($x, y, z, t$) and model the full set of thermodynamical parameters. In addition, this method has the potential to efficiently address level populations and radiative transfer, which are primary sources of long compute times for recent methods \citep{SocasNavarro_2000,ruiz2022desire}.


\section{Data availability}
\label{sec:dataavail}
All our inversion results are publicly available.
\begin{itemize}
    \item Data: \url{https://app.globus.org/file-manager?origin_id=aa4a093a-5b00-4a4b-b8d4-5bb65f324c8c&origin_path=%2F}
    \item Code: \url{https://github.com/RobertJaro/pinn-me}
\end{itemize}


\section{Acknowledgments}
RJ was supported by the NASA Jack-Eddy Fellowship. MEM was supported through a UCAR/ASP Postdoctoral Fellowship.
This material is based upon work supported by the NSF National Center for Atmospheric Research (NCAR), which is a major facility sponsored by the U.S. National Science Foundation under Cooperative Agreement No. 1852977. We would like to acknowledge high-performance computing support from the Derecho 
system~\citep{Derecho_HPC_System} provided by NCAR, sponsored by the National Science Foundation.
We would like to thank Dr. Ivan Milić for providing the synthetic stokes profiles of the MURaM snapshot. 

This research has made use of SunPy \citep{sunpysoftware2020, sunpycommunity2020}, AstroPy \citep{2013A&A...558A..33A} and PyTorch \citep{pytorch2019_9015}.


%

\vspace{5mm}
\facilities{Hinode/SOTSP}


\software{astropy; numpy; matplotlib; pytorch; sunpy}





\bibliography{sample631}{}
\bibliographystyle{aasjournal}



\end{document}